\documentclass{article}

\usepackage{arxiv}

\usepackage[utf8]{inputenc} % allow utf-8 input
\usepackage[T1]{fontenc}    % use 8-bit T1 fonts
\usepackage{hyperref}       % hyperlinks
\usepackage{url}            % simple URL typesetting
\usepackage{booktabs}       % professional-quality tables
\usepackage{amsfonts}       % blackboard math symbols
\usepackage{nicefrac}       % compact symbols for 1/2, etc.
\usepackage{microtype}      % microtypography
\usepackage{lipsum}

\usepackage{amsmath,amssymb}
\usepackage{subfigure} % subfiguras
\usepackage[]{algorithm2e}
\usepackage{float}
\usepackage{graphicx,xcolor}

\title{Driver fatigue EEG signals detection by using \\ robust univariate analysis}

\author{
  Antonio Quintero-Rinc\'on, Carlos D'Giano \\
  Fundaci\'on Lucha contra las Enfermedades Neurol\'ogicas Infantiles (FLENI)\\
  Buenos Aires, Argentina \\
  \texttt{tonioquintero@ieee.org} \\
   \And
 Mar\'ia Eugenia Fontecha \\
  Deparment of Bioengineering, Instituto Tecnol\'ogico de Buenos Aires, Argentina}

\begin{document}
\maketitle

\begin{abstract}
Driver fatigue is a major cause of traffic accidents and the electroencephalogram (EEG) is considered one of the most reliable predictors of fatigue. This paper proposes a novel, simple and fast method for driver fatigue detection that can be implemented in real-time systems by using a single-channel on the scalp.  The method based on the robust univariate analysis of EEG signals is  composed of two stages.  First, the most significant channel from EEG raw is selected according to the maximum variance. In the second stage, this single channel will  be used to detect the fatigue EEG signal by extracting four feature parameters.  Two parameters estimated from the robust univariate analysis, namely mean and covariance, and two classical statistics parameters such as variance and covariance that help to tune the robust analysis. Next, an ensemble bagged decision trees classifier is used in order to discriminate fatigue signals from alert signals. The proposed algorithm is demonstrated on 24 EEG signals from the Jiangxi University of Technology database using only the most significant channel found, which is located in the left tempo-parietal region where spatial awareness and visual-spatial navigation are shared, in terms of 92.7\% accuracy with 1.8 seconds of time delay.
\end{abstract}

% keywords can be removed

	\section{Introduction}
	%\IEEEPARstart{T}{he} 
	
	The electroencephalogram (EEG) is a recording of the electrical activity of the brain, detected by electrodes placed on the scalp. Variations in the brain waves traced by the EEG are correlated with neurological conditions, physiological states and level of consciousness\cite{Mosby2013}.
	
	Driver fatigue is a major cause of road accidents, as 20\% of traffic accidents are related to fatigue driving\cite{Balasubramanian2018}. These accidents cause greater driver mortality and morbidity than other types of accident because of the greater speed on impact \cite{Horne1995}.  In order to prevent this kind of accidents from happening, real-time driver fatigue detection is of great importance as it allows the development of devices capable of warning drivers who are showing signs of fatigue, especially those driving in monotonous conditions.
	According to Lal and Craig \cite{Lal2002}, EEG may be one of the most reliable predictors of driver fatigue. Numerous researchers have, hence, studied different methods based on EEG signals analysis for the detection of driver fatigue.
	
	When analyzing an EEG signal on its frequency domain, four components or bands can be obtained: delta ($\delta$) (0-4 Hz), theta ($\theta$) (4-8 Hz), alpha ($\alpha$) (8-13 Hz) and beta ($\beta$) (13-20 Hz) \cite{Jap2009}. Lal et al. \cite{Lal2002} discovered that delta and theta magnitude increased by 22\% and 26\%, respectively, during the transition to the fatigue state and that alpha and beta activity increased by 9\% and 5\%, respectively. Based on these results, Lal and Craig \cite{Lal2003} then developed a reliable fatigue-detecting algorithm to categorize the mentioned frequency bands into four phases: alert, transition to fatigue, transitional-posttransitional, and posttransitional. Jap et al. \cite{Jap2009} assessed the four activities during monotonous driving and performed four algorithms, which showed an increase in the ratio of slow-wave to fast-wave EEG activities; the algorithm $\frac{\theta + \alpha}{\beta}$ showed a larger increase. Wei et al. \cite{Wei2012} also transformed the collected EEG data into bands ($\theta, \alpha$ and $\beta$) and determined the optimal indicator of driver fatigue out of twelve types of energy parameters; then, using Kernel Principal Component Analysis (KPCA) FP1 and O1 were selected as the significant electrodes. Finally, the model was evaluated and the results showed an accuracy of 92.3\%. Simon et al. \cite{Simon2011} studied the advantages of alpha spindles (short (0.5-2 s) bursts of high-frequency alpha activity \cite{Lawhern2013}) compared to band power measures as a fatigue indicator and found that alpha spindle parameters increase both fatigue detection sensitivity and specificity.
	
	Other authors have applied entropy methods to study the detection of driver fatigue on EEG signals. Hu \cite{Hu2017} employed four types of entropy measures (spectral entropy, approximate entropy, sample entropy, and fuzzy entropy) to extract features from a single EEG channel and then compared them by ten classifiers. The optimal performance was achieved when combining channel CP4, fuzzy entropy and Random Forest classifier; the accuracy was of 96.6\%.  Mu et al. \cite{Mu2017} used the four types of entropy measures combined to extract features from EEG signals (recorded in both alert and fatigue states). In terms of fatigue detection, results showed that combined entropy has superior performance compared to single entropy, with the average recognition accuracy being 98.75\%. Performance of the electrodes was also assessed and results showed that T5, TP7, TP8, and FP1 performed better. Min et al. \cite{Min2017} proposed a multiple entropy fusion analysis to detect driver fatigue in an EEG-based system and achieved an accuracy of 98.3\%, a sensitivity of 98.3\% and a specificity of 98.2\%. The study of electrode performance concluded that the significant electrodes were T6, P3, TP7, O1, Oz, T4, T5, FCz, FC3 and CP3. See \cite{Elassad2020} for  a complete state-of-the-art about the application of machine learning techniques for driving behavior analysis.
	
	In a statistical context, the multivariate and univariate analysis are used to analyze possibly correlated data containing observations, an EEG signal in our case. By the other hand, the robust analysis consists of techniques that are robust to non-normality or to outliers. Note that the classical mean and covariance methods are very robust to non-normality, but are not robust to outliers \cite{Olive2017}. See \cite{Molinari1991,Yong2008,Sameni2017,Uehara2017} for some works in EEG signal processing using robust analysis. Also, an interesting topic is that the EEG signals have the assumption that the observed data has a Gaussian distribution \cite{QuinteroRincon2018a}, which permits an approximate representation to the EEG real data sets, and at the same time it is theoretically possible to apply the optimal statistical methods\cite{Maronna2019}.
	
	In neuroscience context, the multivariate EEG signals are related to the state of many regions simultaneously as well as their interactions, while the univariate analyze a single brain region \cite{QuinteroRincon2016b}.
	
	The aim of this study was to propose a novel, simple, and fast method for driver EEG fatigue signals detection that can be implemented in real time by using a single-channel on the scalp. The method based on the robust univariate analysis of EEG signals was composed of two stages.  First, the most significant channel from EEG raw was selected according to the maximum variance. In the second stage, this single-channel will  be used to detect the fatigue EEG signal by extracting four features parameters.  Two parameters estimated from the robust univariate analysis, namely mean and covariance, and two classical statistics parameters such as variance and covariance that help to tune the robust analysis for driver fatigue EEG signal detection that can be implemented in real time. This motivation arose for two reasons. The first is related to the level of traffic accidents due to driver fatigue and the other one to the interesting problem of studying a single brain region that has a relation with the other brain regions. In other words, a univariate analysis from the multivariate situation. Remember that, in multivariate EEG signals the information for each channel is considered in relation to the other channels. Additionally it is normal that in the majority of the observations, some channel values fail to maintain the pattern of relationships between the channels evident. Therefore, robust univariate analysis is an attractive approach to this topic. 
	
	In this work, we analyzed  32-channel fatigue EEG multivariate signals from 12 subjects from the database from the  Jiangxi University of Technology \cite{Min2017}, in two stages. In the first stage, for each channel of all subjects, the variance was computed in order to determine a single-channel to be able to estimate the changes between alert to fatigue during long driving times.  The most significant channel in this database was a channel in the left tempo-parietal region. This region is very important because it involves spatial awareness and visual-spatial navigation. Next, in the second stage, four feature parameters were computed from this univariate signal as follows: two parameters estimated from the robust univariate analysis, namely mean and covariance, and two classical statistics parameters such as variance and covariance. Therefore, our goal was to estimate a univariate EEG region that allowed the detection of an EEG fatigue signal by using robust univariate analysis coupled with two classical statistics methods. Note that all these techniques are well known; however, to the best of our knowledge, this promising method has not been investigated for EEG signal fatigue detection and it is even potentially interesting for real time implementation. Next, a supervised classification method through the ensemble bagged decision trees was used in order to discriminate fatigue signals from alert signals. Precisely, these two sets of vectors, each one with the four parameters explained before are used for testing, training and prediction stages. Due to the high dynamic variability of the EEG signals, this classifier permits increasing the size of the training set, decreasing the variance, and increasing the accuracy and narrowly tuning the prediction to the expected outcome \cite{QuinteroRincon2019a}.
	
	The remaining paper is organized as follows. Section \ref{sec:lob} explains the importance of the lobes of the brain and their relation with the EEG electrodes distribution across the scalp. Section \ref{sec:mm}
	describes the database used and the methodology proposed, then in Section \ref{sec:res} the methodology is applied, analyzed and discussed in real EEG signals from the database from Jiangxi University of Technology \cite{Min2017}. Finally, in Section \ref{sec:con} the conclusions and future works are presented.
	
	\section{Lobes of the brain}
	\label{sec:lob}
\begin{figure}[!h]
		\centering
		\subfigure[Lobes of the brain, figure adapted from \cite{lumen2019}]{\includegraphics[clip,width=0.45\columnwidth]{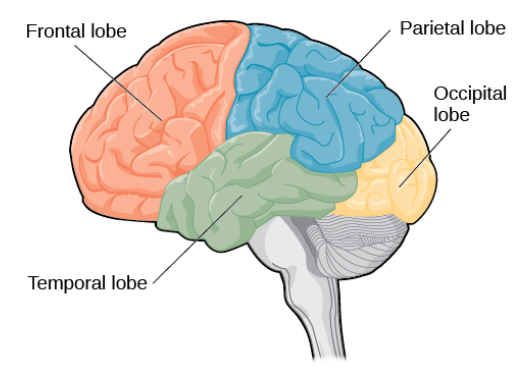}}\label{fig:lob}
		\subfigure[EEG electrodes positions using the 10-20 international system]{\includegraphics[clip,width=0.4\columnwidth]{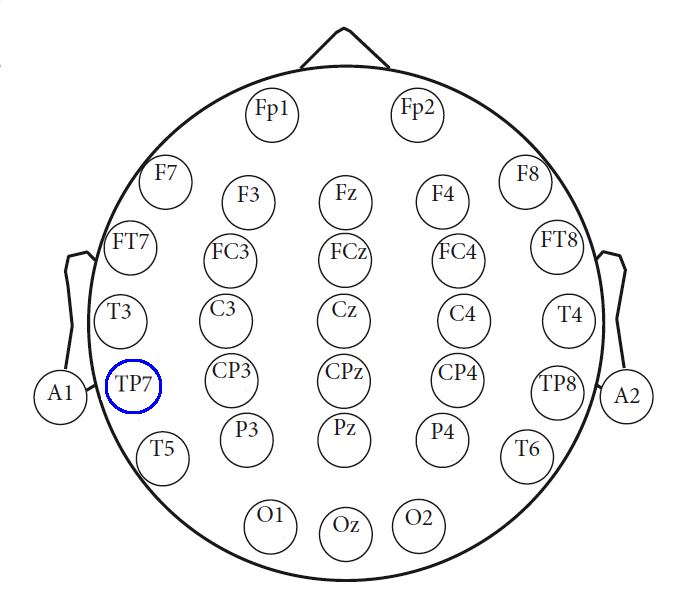}}\label{fig:eeg}
		\caption{The figures show how the lobes of the brain can be correlated with the positions of the electrodes according to their nomenclature,  (T) temporal lobe, (P) parietal lobe, (O) occipital lobe and (F) frontal lobe.}
		\label{fig:descriptions}
	\end{figure}

	The brain's main task is to help keep the whole body in an optimal state in relation to the environment in order to maximize survival opportunities. During this process, the brain classifies the most relevant information as electrical impulses from neurons in the sense organs. This electrical activity is amplified causing them to be represented in various brain regions through the different lobes namely frontal, parietal, occipital and temporal. If this activity is sustained for long enough, it will result in a conscious experience \cite{Carter2019}. Figure \ref{fig:descriptions} shows the different lobes of the brain and the EEG electrodes positions that reference each lobe according to the 10-20 international system. The electrodes are labeled according to their lobe location and hemisphere involved, e.g. F = frontal, P = parietal, O = occipital, T = temporal, C = central, odd numbers for left, even numbers for right, and z for the midline.
	
	The main characteristics of the frontal lobe are motion control, reasoning,  conscious emotion, language, planning, thinking, making judgments, and decision-making. The parietal lobe is related to the integration of sensory and perceptual stimuli both consciously and unconsciously such as spatial computation, body orientation, and attention. The temporal lobe is associated with memory, hearing, language, emotion and visual-spatial navigation. The occipital lobe integrates all the information that we receive visually both consciously and unconsciously, in other words, it integrates the visual processing. The central groove splits the frontal and the parietal lobes and it is related to the primary motor cortex and the somatosensory cortex. The lateral groove separates the frontal and parietal lobes from the temporal lobe, it is related to language production. See \cite{Niedermeyer2011,Neuroscience2012} in order to expand this information.
	
	\section{Material and Method}
	\label{sec:mm}
	
	\subsection{Database}
	\label{ssec:db}
	A database with 12 unipolar fatigue EEG signals and 12 unipolar alert EEG signals were used from \cite{Min2017}. In this experiment, twelve university male students with ages between 19 and 24 years were asked to drive a static car driving simulator in a software-controlled environment. For each subject, two EEG signals were recorded: one corresponding to alert driving and the other one corresponding to fatigue driving. The alert driving EEG signal was recorded after the subject had been driving for 15 minutes. The fatigue driving EEG signal was acquired once the subject showed signs of fatigue according to Lee's subjective fatigue scale and Borg's CR-10 scale \cite{Borg1990} \cite{Lee1991}, after driving for 60-120 minutes. Both signals consist of a 32-channel EEG, have a duration of 5 minutes and were digitized at 1000 Hz.
	
	\subsection{Methodology}
	\label{ssec:meth}
	The proposed methodology was composed of two feature extraction stages. The first feature extraction stage is related to the best channel estimation in order to detect signal fatigue in EEG signals from the database introduced above in \ref{ssec:db}. The second feature extraction is for using this channel to classify, detect and predict fatigue in EEG signals using the robust univariate analysis coupled with the variance and the covariance. Note that, these two classical statistics methods do not introduce any additional computational cost. We now introduce the nomenclature used in this paper.
	
	Let $\boldsymbol{X} \in \mathbb{R}^{M\times N}$ denote the multivariate matrix gathering $M$ EEG signals $\boldsymbol{x}_m \in \mathbb{R}^{1\times N}$ measured simultaneously on different channels and at $N$ time instants. To each channel $M$ the variance ($\sigma^2$) is computed in order to estimate the channel with the maximum variance by using
	\begin{align}
	\label{eq:var}
	\sigma^2= \max_{\left[\boldsymbol X\right]^T}  \left[ \frac{1}{N-1}\sum_{i=1}^{N}\left| \boldsymbol x_{m,i}-\mu \right|^2
	\right] 
	\end{align}
	where $1 \leq m \leq M$, and $\mu$ is the mean of $\boldsymbol x_{m}$, so that
	\begin{align}
	\label{eq:mean}
	\mu= \frac{1}{N}\sum_{i=1}^N \boldsymbol x_{m,i}
	\end{align}
	Let $\Theta=\left[\theta_1,\cdots,\theta_N\right]^T$ the selected univariate channel in order to detect the fatigue signal in EEG signals by using four features. The first two features are related to the robust analysis through the mean of $\Theta$ for which the determinant of the sample covariance is minimal ($\widehat{\mu}$), and the corresponding covariance matrix multiplied by a consistency factor $c_0$ ($\widehat{\Sigma}=c_0\Sigma$), given by
	\begin{align}
	c_0 = \frac{\alpha}{F_{\Theta^2_{p+2}}\left(\Theta^2_{p,\alpha} \right)}
	\end{align}
	where $p=1$ is the univariate channel, $\Theta^2_{p,\alpha}$ is the $\alpha$-quantil (97.5\%) of the $\Theta^2_{p}$ distribution, with $\alpha=\lim_{n\to \infty} h(n)/n$, where the minimum covariance determinant estimator must meet $[n+p+1]\leq h\leq n$, see \cite{Hubert2010,Rousseeuw1999} for more details.
	
	The other two feature parameters are the variance and the covariance ($\Sigma$) using 
	\begin{align}
	\label{eq:cov}
	\Sigma(\Theta,\Theta)= \frac{1}{N}\sum_{i=1}{N}(\Theta-\mu_{\Theta})^*(\Theta-\mu_{\Theta})
	\end{align}
	where $*$ denote the complex conjugate and $\mu_{\Theta}$ is the mean. Note, that the two classical statistics features help to tune the robust analysis and they do not introduce any additional computational cost.

	Each feature was estimated splitting the univariate channel $\Theta$ into 2-second time segments using a rectangular sliding window with 0.5-second time overlapping. Finally, every feature was normalized between $-1$ and $1$ for all each 2-second time segments \cite{QuinteroRincon2019d}. 
	Therefore, we have two  sets  of  vectors each  one  with the 4 features introduced above.  $\Lambda_1=[\widehat{\Sigma},\widehat{\mu},\sigma^2,\Sigma]$
	for fatigue EEG signals and $\Lambda_0=[\widehat{\Sigma},\widehat{\mu},\sigma^2,\Sigma]$ for alert EEG signals. These two sets of vectors were estimated in all database in order to classify between fatigue or alert EEG signals by using ensemble bagged trees classifier. This classifier is a technique used to reduce the variance of the predictions $\Lambda_1$ and $\Lambda_2$, by combining the result of multiple classifiers modeled on different sub-samples through decision trees of the same dataset $\mathcal{D}$, see \cite{Flach2012,QuinteroRincon2019a} for more details.
	
	Let $\mathcal{D}=\{d_1,\cdots,d_N\}$ with $d_i=(\lambda_i,c_i)$, where $c_i$ is a class label, $1$ for fatigue EEG signals and $0$ for alert EEG signals. The ensemble bagged decision trees classifier returns the ensemble as a set of models through the Algorithm \ref{alg:bag}.\\
	
	\begin{algorithm}[!h]
		\KwData{data set $\mathcal{D}$; ensemble size $\mathcal{T}$; learning algorithm using decision trees $\mathcal{A}$}
		\KwResult{ensemble of models whose predictions are to be combined by majority vote.}
		\For{t=1 to $\mathcal{T}$}{
			build a bootstrap sample $\mathcal{D}t$ from $\mathcal{D}$ by sampling $|\mathcal{D}$ data points with replacement\;
			run $\mathcal{A}$ on $\mathcal{D}t$ to produce a model $\mathcal{M}_t$\;
		}
		\caption{Bagged($\mathcal{D}$,$\mathcal{T}$,$\mathcal{A}$) - train an ensemble of models from bootstrap samples, adapted from \cite{Flach2012}}
		\label{alg:bag}
	\end{algorithm}
	
	\section{Results and discussion}
	\label{sec:res}
	In this section we evaluate the proposed methodology using the database introduced in subsection \ref{ssec:db}.  
	
\begin{figure}[!ht]
		\centering
		\includegraphics[clip,width=1.0\columnwidth]{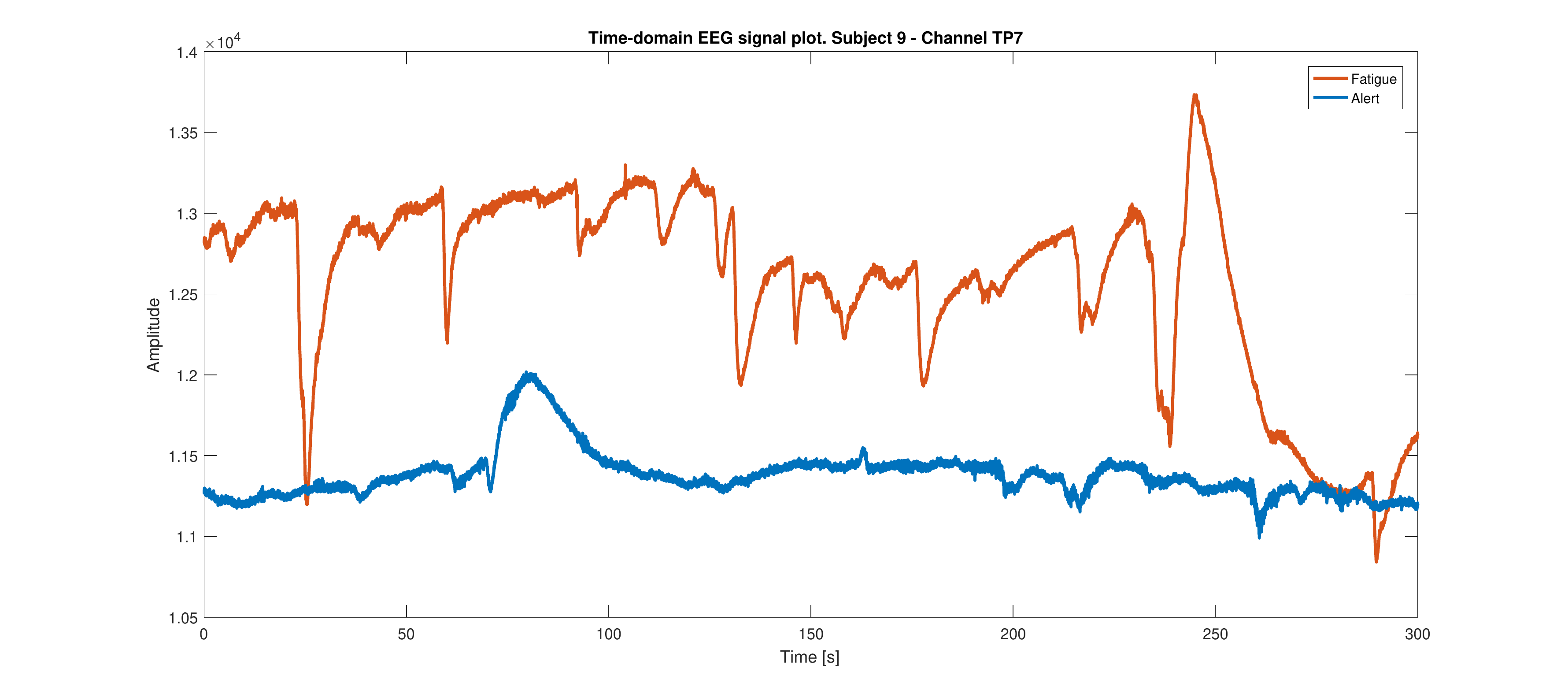}
		\caption{Alert (blue) and fatigue (orange) EEG signals plot. Fatigue signal shows higher variability than the alert signal.}
		\label{fig:seniales}
	\end{figure}

	Below is an example of the signals to which the variance was calculated to determine the most significant electrode. Figure \ref{fig:seniales} shows a time-domain plot of both alert and fatigue EEG signals corresponding to electrode TP7, subject 9, obtained from the database. 
	
\begin{figure}[!ht]
		\centering
		\includegraphics[clip,width=1.0\columnwidth]{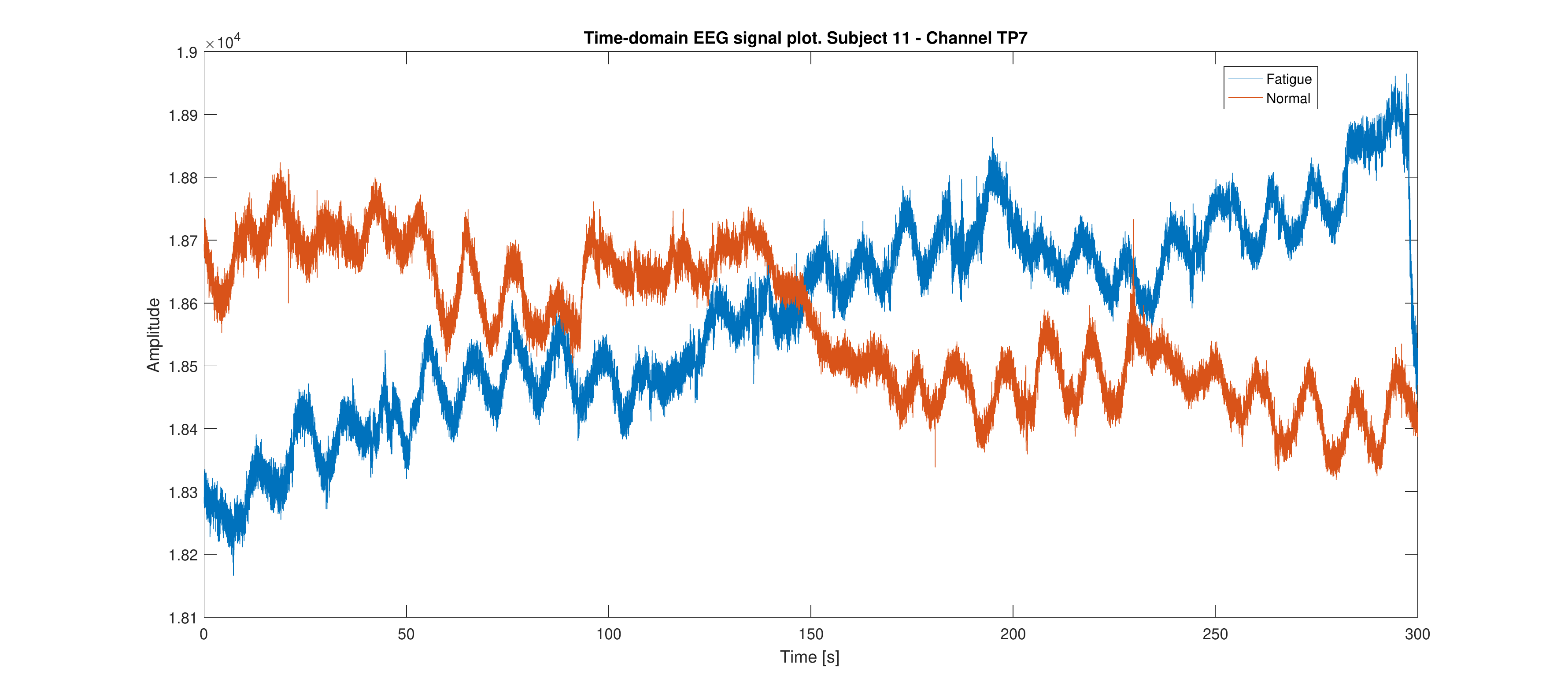}
		\caption{Alert (blue) and fatigue (orange) EEG signals plot. Fatigue and alert signals amplitudes don't show significant differences.}
		\label{fig:seniales2}
	\end{figure}

	By visually analyzing the signals it can be clearly seen that fatigue signal (orange) shows a higher variability compared to the alert signal (blue). Note that, in this example, it is easy to differentiate the amplitude of the signals, but in general due to the high dynamic of the EEG signals, it is not always that way. Figure \ref{fig:seniales2} shows an example of this.
	
	We hypothesize that the melatonin is responsible for the high variability temporal waveform of the signal, see Figure \ref{fig:seniales}. It is a hormone produced by the pineal gland that helps regulate the sleep-wake cycle. According to \cite{Carter2019}, the suprachiasmatic nucleus (SCN) in the hypothalamus plays a key role in sleep-wake cycles. Light levels are sensed by the retina, and this information is relayed to the SCN, which then sends a signal to the pineal gland. This triggers the release of melatonin, the hormone that tells the body when to sleep. At this point, the brain becomes less alert and fatigue starts to take over. When melatonin levels fall in response to increased light, the waking part of the cycle begins.
	
	The best channel by using the equation \eqref{eq:var} from subsection \ref{ssec:meth} in all database, was TP7, a channel in the left tempo-parietal region where the spatial awareness and visual-spatial navigation are shared, see Figure \ref{fig:eeg} from Section \ref{sec:lob}. Besides, the physical structure of the brain reflects its mental organization. In general, higher mental processes occur in the upper regions, while the brain's lower regions take care of basic support \cite{Carter2019}. Therefore, this finding is promising since is related to the cautiousness, an affective faculty \cite{Postle2015}. 
	
	To test the quality of our fatigue EEG signal detection an ensemble bagged trees classifier was used. The classifier was trained and tested off-line with 10-fold cross-validation technique with the two sets of vectors with the 4 features estimated in subsection \ref{ssec:meth} for each vector $\Lambda_1=[\widehat{\Sigma},\widehat{\mu},\sigma^2,\Sigma]$
	for fatigue EEG signals and $\Lambda_0=[\widehat{\Sigma},\widehat{\mu},\sigma^2,\Sigma]$ for alert EEG signals. It is important to clarify that our data are balanced, therefore, the classifier used do not have minority classes to be overlooked \cite{ImbalancedData2018,QuinteroRincon2019e}.
	To evaluate the performance of a proposed method different metrics may be used. Elassad et al. \cite{Elassad2020} is a recent study that analyzed the performance metrics used in different studies about driving behavior and found that accuracy, recall and specificity were the most popular ones. According to this study, the \emph{accuracy} was adopted in 65.85\% of the 82 primary studies published in the decade between 2009 and 2019, while recall and specificity were used in 35.36\% and 32.92\% of them, respectively. We adopt accuracy as the main measure of performance, although we also inform the sensitivity and specificity of our method.
	 
	 The percentage of correct classifications was
	analyzed, in terms of sensitivity, specificity and accuracy. The values obtained are: 91\% sensitivity (True positive rate), 95\% specificity (True negative rate), 92.7\% accuracy and 97\% area under curve for fatigue detection in EEG signals. For each iteration, 14450 observations were used obtaining a prediction with a time delay of 1.8 seconds, and $p$-value $< 0.01$ using ANOVA test.
	
	 Table \ref{tab:channels} reports some examples of different channels used in different works. For example, in \cite{Wei2012} the Grey relational analysis (GRA) and the Kernel  Principal  Component  Analysis (KPCA) were used in order to extract the more principal channels and then the model was evaluated with a linear regression equation (LRE); only the accuracy is reported. Hu
	\cite{Hu2017} compared the effectiveness of four different features combined with ten different classifiers to detect driver fatigue and concluded that optimal performance was achieved using fuzzy entropy (FE) features and Random Forest (RF) classifier, obtaining an accuracy of 96.6\%. In \cite{Mu2017} a combined entropy-based method with the support vector machine (SVM) classifier showed 98.75\% accuracy, 97.50 \% specificity and 96\% sensitivity. Min et al. \cite{Min2017} proposed a multiple entropy fusion (EF) method and when evaluating the model with a back propagation neural network classifier (BP) the accuracy obtained was 98.3\%. Yeo et al. \cite{Yeo2009} used power spectrum features from 19 channels and SVM to distinguish between alert and fatigue EEG signals. The model achived an accuracy of 99.3\%. In \cite{Chen2018} Chen et al. perfomed the fusion of two features: the fuctional brain network (FBN)-based feature and the fatigue PSD-based feature. Integrating the fusion feature (FBN-PSD-FF) with an extreme learning machine (ELM) classifier, they developed an automated fatigue detection system that showed an accuracy of 95\%, a sensitivity of 95.71\% and a specificity of 94.29\%.
	
	\begin{table}[htbp]
		\caption{Different channels used in different works. In general, all of them present a high accuracy (Acc.). RUA = Robust univariate analisys, EBDT = Ensemble bagged decision trees, GRA = Grey relational analysis, KPCA = Kernel  Principal  Component  Analysis, RF = Random Forest, SE = Sample entropy , FE = Fuzzy entropy, AE = Approximate entropy, PE = Spectral entropy, SVM = Support vector machine, EF = Entropy fusion, BP = Back propagation, FBN = Functional brain network, PSD = Power spectral density, ELM = Extreme learning machine. }
		\label{tab:channels}
		\centering
		%\begin{tabular}{ | p{2cm} | p{1.4cm} | p{1.4cm} |p{0.7cm} | p{0.4cm} |}

		\begin{tabular}{|l|l|l|l|l|}
			\hline \hline
			Channels &  Methods & Classifier & Acc. & Ref. \\
			\hline \hline
			TP7 & RUA & EBDT & 92.7\%& our \\ \hline
			FP1, O1 & GRA + KPCA & LRE & 92.3\% & \cite {Wei2012} \\ \hline
			CP4 & SE + FE + AE + PE & RF & 96.6\% &\cite{Hu2017} \\ \hline
			T5, TP7, TP8, FP1 & SE + AE + SE + FE & SVM& 98.75\% & \cite{Mu2017} \\ \hline
			T6, P3, TP7, O1, Oz, T4, T5, FCz, FC3, CP3 & EF & BP & 98.3\% & \cite {Min2017} \\ \hline
			All channels & PSD & SVM & 99.3\% & \cite {Yeo2009} \\ \hline
			All channels & FBN-PSD & ELM &95.0\% & \cite {Chen2018} \\ \hline
			\hline
		\end{tabular}
	\end{table}
	
	Note that, although the methods have good accuracy, in general, they are complex and have a high computational cost. Therefore we suggest that our method is computationally very efficient and has advantages for real-time implementation systems by using a single-channel on the scalp.

	\section{Conclusions}
	\label{sec:con}
	This work presented a new method to detect driver fatigue events in EEG signals using one single-channel within  the left tempo-parietal region where spatial awareness  and  visual-spatial  navigation  are  shared, one important contribution.
	The method is based on the robust univariate analysis of EEG signals that captures the EEG signal through two feature parameters namely mean and covariance. Added to this, other two parameters are estimated using classical statistics methods such as variance and covariance that help to tune the robust analysis. In total, four features were estimated through two sets of vectors: one for fatigue EEG signals, and the other one for alert EEG signals. Next, the ensemble bagged classifier was used with these two sets of vectors in order to discriminate fatigue from alert events in EEG signals and obtained an accuracy of 92.7\%. The proposed method was demonstrated on a real database from Jiangxi University of Technology \cite{Min2017}, containing 24 multivariate EEG signals. 
	
	This method had two main limitations. The first one was to define a perfect sliding time-window with perfect overlap of the segments, and the second one was that the classical mean and covariance methods are not robust to outliers. On the other hand, the main advantage is that the method can be implemented in real time due to its simplicity by using a single-channel on the scalp.
	
	Future work will focus on the optimization of the method in order to be applied to other EEG signals with high inter-variability, such as epileptic signals that are very susceptible to outliers, masking or swamping \cite{Rousseeuw2011}.  Another important topic to explore is about biased competition \cite{Landau2018}, which is related to the attention that causes neurons to tune in to stimuli that are relevant, over other, irrelevant stimuli.

	\bibliographystyle{IEEEtran}
	%\bibliography{biblio}

\end{document}